# Laser-heating-based active optics for synchrotron radiation applications

FUGUI YANG*, MING LI, XIAOWEI ZHANG

*Laboratory of X-ray Optics and Technologies, Beijing Synchrotron Radiation Facility, Institute of High Energy Physics, Chinese Academy of Sciences, Beijing 100049, Beijing, China*
*Corresponding author: yangfg@ihep.ac.cn*

**Active optics has attracted considerable interest from researchers in synchrotron radiation facilities, because of its capacity for x-ray wavefront correction. Here, we report a novel and efficient technique for correcting or modulating a mirror surface profile based on laser-heating-induced thermal expansion. An experimental study of the characteristics of the surface thermal deformation response indicates that the power of a milliwatt laser yields a bump height as low as sub-nanometer scale, and that variation of the spot size modulates the response function width effectively. In addition, the capacity of the laser-heating technique for free-form surface modulation is demonstrated via a surface correction experiment. The developed method is a promising new approach towards effective x-ray active optics coupled with at-wavelength metrology techniques. © 2016 Optical Society of America**

http://dx.doi.org/10.1364/OL.99.099999

Modern third-generation synchrotron radiation sources and x-ray free electron lasers (FELs) can generate highly coherent and bright x-ray beams[1,2]. In order to successfully exploit the full potential of such sources for nanoscale matter research, which requires preservation of the coherence and realization of the diffraction limit, the x-ray optics used in synchrotron beamlines must be refined. Improvement of the x-ray optics of a beamline, which transports and modulates a synchrotron beam, is challenging in terms of both manufacturing and metrology techniques.

The use of a high-quality mirror produced using modern manufacturing techniques, for example, via the "super-polishing" method reported by Yamauchi et al. [3], can improve the slope error to only tens of nano radians for a plane mirror. However, it is still difficult to build a satisfactory x-ray optics system with highly curved surfaces, because of the performance limitations of current metrology equipment. Nonetheless, even if such state-of-the-art optics technology were employed, the wavefront aberration could be distorted as a result of imperfect upstream optics (the monochromator, windows, etc.) or transmission window errors due to the photon heat load, mechanical clamping distortion, or misalignments. As regards the thermal bump in the synchrotron radiation beamline [4], for example, an experiment conducted by Rutishauser et al. [5] using x-ray grating interferometry has shown that the monochromator heat bump is of the order of 20 nm for a 1-W incident x-ray power.

Recently, active optics, or deformable mirrors, which include piezo bimorph mirrors [6, 7] and mechanically bendable mirrors [8], have been widely studied. These mirrors are expected to facilitate focal-length controllability or wavefront error compensation. Thus, the use of active optics can reduce the requirement for high shape accuracy for x-ray mirrors. For example, 7-nm nano-focusing has been successfully achieved using Kirkpatrick-Baez mirrors in combination with bimorph mirrors. To date, x-ray adaptive optical systems have been employed as standard tools in many beamlines[9]. However, we note that the shape stability remains problematic. In addition, because of the restricted space for the mechanical mounts of the actuators, it is difficult to increase the number of actuators along the clear aperture of a mirror in order to correct surface errors with high spatial frequency. In some experiments, this may induce problems.

Such problems may be overcome using laser heating. Thermal stress is not a new research subject and has been widely studied in many engineering fields, especially in relation to laser-induced heat, because of its easy modulation. One example is the wide use of laser-heating techniques in industrial materials processing, e.g., for laser ablation [10] and nano manufacturing [11]. In these applications, extremely high temperatures approaching the boiling point are generally employed to etch or melt solid matter. In other applications, however, the thermal effect has a deleterious influence on the system performance and is regarded as a serious problem, e.g., thermal lensing in continuous-wave (CW) end-pumped solid-state lasers [12, 13]. Many theoretical studies involving both full numerical solutions and analytic approximations also have been conducted to determine methods of managing positive or negative laser-heating effects efficiently [14, 15].

In this Letter, we propose a laser-heating-based technique to deform the shape of a synchrotron mirror freely and easily, which is expected to be employed as a wavefront correction method. To the best of our knowledge, the use of laser-heating effects to shape mirror surfaces in synchrotron facilities has not been reported previously. In order to demonstrate this technique, CW lasers at visible wavelength are used to illuminate and heat a cuboid mirror, the details of which are given below. The surface error is corrected efficiently (being reduced from

10 to 3 nm root mean square (RMS)), following easy adjustment of the laser power. Note that, because of the low power requirement for nanometer (nm)-level surface correction, this method can be realized using low-cost laser diodes (LDs). Moreover, the good spatial and temporal modulation properties of the lasers facilitate easy adjustment of the width and height of the heat bump; therefore, this deformation method can correct wavefront/surface errors over a wide spatial wavelength range. We hope that the results presented in this Letter can stimulate new investigation in this field.

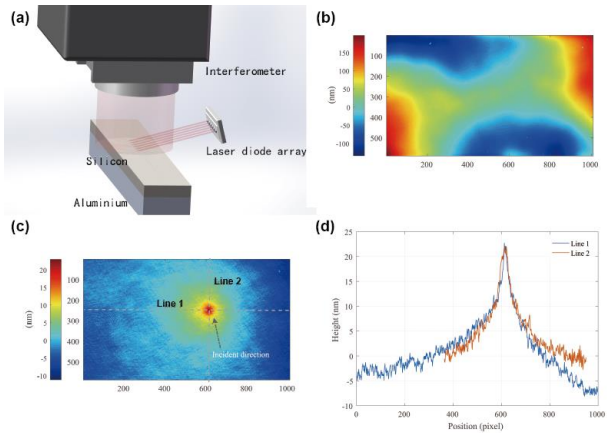

Fig. 1. (a) Schematic of experimental setup used to test x-ray mirror deformation under laser-diode (LD) laser heating. (b) Mirror surface profile without laser irradiation. PV and root mean square (RMS) surface error values are 250 and 10.06 nm, respectively. (c) The change of the surface height profile after laser-heating (incident laser power ~ 300 mW). . (d) Profile plots along line1 and 2 marked in (c) showing the bump with 20-nm peak-to-value (PV) value.

The experimental setup is illustrated in Fig. 1(a). The cuboid mirror was comprised of an uncoated silicon crystal with 20-mm thickness and an uncooled aluminum block with 40-mm thickness. The area of the top mirror surface was 500 mm (L) × 80 mm (W). The entire block was placed horizontally on a six-degree-of-freedom sample stage. Above the mirror, a Fizeau interferometer (Zygo, Verifire-QPZ™) based on the phase-shifting interferometry (PSI) technique was used to characterize the surface shape precisely. A laser-beam array was delivered directly onto the mirror surface. Each beam was produced by one LD, the stability and maximum power of which were ~5% and 500 mW, respectively. A PM100 laser power meter with an S310 standard thermal sensor (Thorlabs, USA) was used to monitor the incident and reflected beams. Because of the smooth surface of the mirror, light loss due to scattering could be neglected, and the heat power absorption was computed by subtracting this component. In order to increase the laser-heating efficiency, the incident angle was set close to the Brewster angle and the fast axis was in the incident plane. However, because of the low degree of polarization of the laser, the reflection loss of the system was somewhat high (approximately 30%) compared to the theoretical value. Note that the other advantage of this arrangement is that it prevented high-powered laser beams from entering the interferometer.

For surface measurement with high accuracy down to 1 nm, a stabilized environment is very important. Here, the results of $N$ set measurements were averaged in order to decrease the random noise. The repeatability of the measurement was found to be approximately 2 nm RMS for $N$ = 50. Figure 1(b) gives the mirror surface profile obtained by eliminating the tilt term. The measurement resolution is 0.125 mm/pixel. The low form error of ~ 300 nm peak-to-value (PV) indicates that the interferometer system error was not significant in this study. In the subsequent laser-heating experiments, we adopted this measurement as a reference in order to determine the laser-heating-induced deformation. Figure 1(c) shows the spatial distribution of a typical surface deformation for 300-mW laser power, while Fig. 1(d) is a plot of the height profiles along the two lines through the peak center marked in Fig. 1(c). Note that the surface-shape stabilization occurred quickly, within a period of less than 1 min. The PV value of the bump was as high as 25 nm and the bandwidth was approximately 40 mm. Both of these results are significant for surface or wavefront correction applications. In addition, the heat bump was also observed to exhibit an asymmetric shape. From the incident laser direction, which is indicated by the arrow in Fig. 1(c), it can be inferred that the narrowness of the mirror, rather than its infinite width causes this asymmetry. The other possible factor, i.e., the oblique incidence of the laser, is excluded, because of the very short absorption depth (0.5 µm) in silicon for a 450-nm laser. In the data processing conducted below, the PV value of the bump was calculated for the same line rather than for the entire surface.

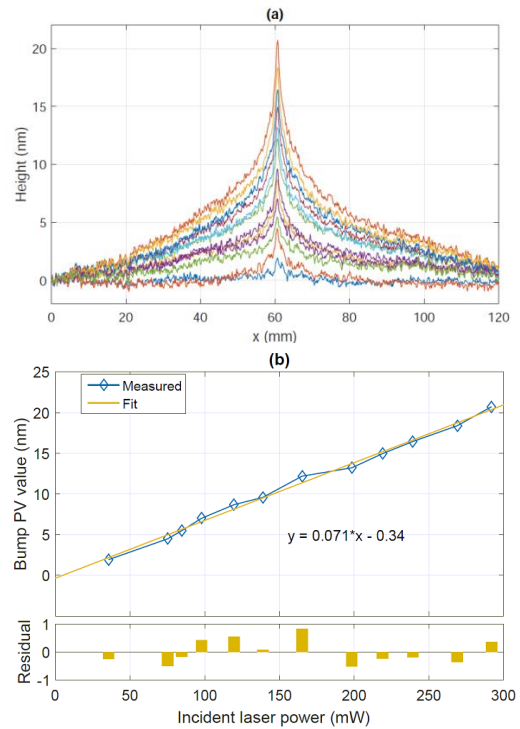

Fig. 2. Surface deformation response to laser heating. (a) Measured surface height profiles for different incident laser powers, and (b) corresponding PV value of deformation as function of laser power.

As regards surface profile correction or active optics, the characteristics of the mirror laser-heating response function are fundamental for application. Therefore, we measured the surface deformation response function considering variation of the laser power. The beam was generated by a diode-pumped solid-state laser (DPSS; maximum power: 300 mW; wavelength: 532 nm). Figure 2(a) plots the deformation height curves along line 1 marked on Fig. 1(c), for incident laser power ranging from 40 to 300 mW. The laser power absorption efficiency was approximately 60%, and the projection area of the focused laser spot was approximately 0.6 mm × 1.3 mm. A heat bump with height as low as 1.9 nm was measured. A plot of the PV value as a function of the laser power and the corresponding linear fitting are shown in Fig. 2(b); the low offset of -0.34 nm verifies the efficacy of the measurement, although the unstable laser source and the error of the interferometer measurement contribute to the overall error. The linear response coefficient is ~0.07 nm/mW, the efficacy of which can be enhanced by increasing the absorption efficiency. Because mature stable lasers of milliwatt (mW)-level power are employed and only a small deformation is required for x-ray wavefront correction, this laser-heating technique can perform very precise surface correction. Note that the other important parameter is the width of the response curve. This value determines the lower spatial wavelength of the corrected surface error. From Fig. 2(a), the $e^{-1}$ bandwidth of the response curve is approximately 6 mm and exhibits small changes with variations in the laser power. This small bandwidth, which is lower than that reported in Ref. [7], indicates the capability of this technique for surface correction with high spatial resolution.

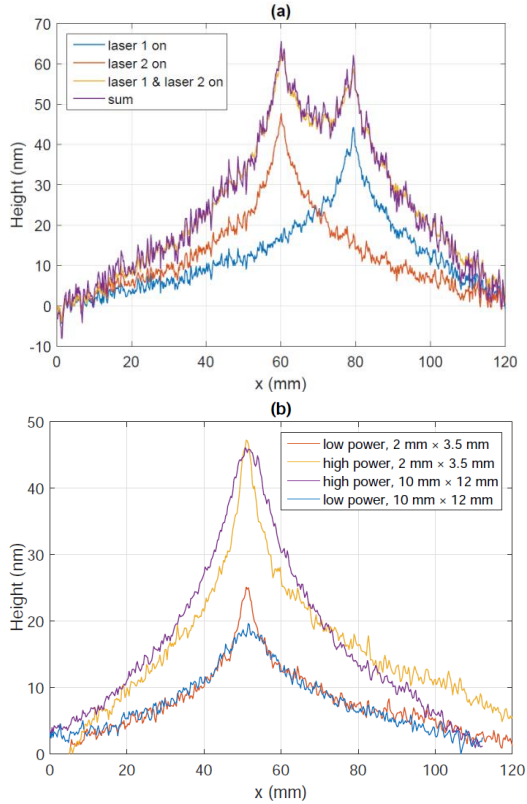

Fig. 3. Spatial superposition effect of laser-heating deformation for (a) two separate laser beams and (b) different beam sizes, where the laser power is adjusted in order to generate deformation with the same PV value.

Apart from the good regulation performance of the laser power, the other advantage of this laser-heating technique is that the response function width can be modulated by the spot size. To confirm this characteristic, the linear superposition effect was examined. Using two high-power LDs (Nichia, NDB7412T; maximum power: 1 W) labeled "1" and "2", we conducted measurements for various laser statuses: 1 on and 2 off; 1 off and 2 on; and 1 on and 2 on. The observed profiles are plotted in Fig. 3(a), where the sum of the two individual curves is also plotted. The perfect match with the measured data demonstrates that the thermal deformation in our case results from linear superposition of the laser beams. The explanation for this behavior is that the laser is transformed into heat on the surface, because of the short absorption depth (< 1 μm). Further, the linear theory of thermoelasticity [16] applies to this planar heat source. In addition, we varied the single spot size and its power, and the results are shown in Fig. 3(b). Here, the response curves correspond to two different projection areas: 2 mm × 3.5 mm and 10 mm × 12 mm. High and low laser powers were considered for each spot. Hence, it is apparent that a larger beam spot generates a large range of deformation, but the required laser power is also high. Overall, these experiments show that both the width and height of the surface deformation can be modulated by changing the laser power and spot size.

Finally, we performed an experiment involving preliminary correction of the surface figure error, in order to confirm that free-form deformation can be achieved via the laser-heating method. In this case, the region of interest was approximately 140 mm smaller than the interferometer aperture. The lasers from an LD array (ML520G1; maximum power: 300 mW; wavelength: 638 nm) with 15-mm pitch were focused and delivered to the surface for correction. The response functions for each laser beam were measured by setting the LDs to the maximum power. Because of the mechanical defects of the lasers in the array, the beams deviated from each other in terms of both spot size and coupling efficiency, resulting in differences in the projection area size and the generated laser power; this is the reason for the significant variations in the response functions shown in Fig. 4(a). However, despite this discrepancy, the realized surface correction is apparent through comparison of the surface height profile of one line before and after the correction process, as shown in Fig. 4(b), especially for the [20 mm, 100 mm] region. A reduction of the figure error from 10.06 to 3 nm RMS was achieved. Referring to Wang [6], the correction can be improved further using a feedback-optimization scheme and using superior lasers. In addition, we note that the residual surface error under short spatial wavelengths is somewhat large, which can be resolved through use of narrower laser beams.

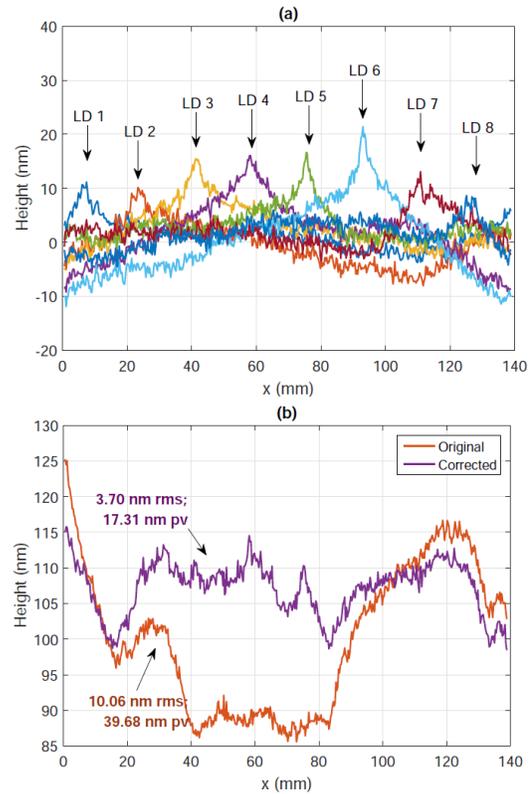

Fig. 4. (a) Laser-heating response function of silicon crystal for eight LDs (LD1–LD8). (b) Correction of mirror surface shape through rough adjustment of laser power.

In conclusion, we have introduced a new active optics method based on the laser-heating effect. Its novelty lies in the use of a commercial laser in order to substantially reduce the complexity and, very likely, the cost of the optomechanical system. Through proper characterization of the surface laser-heating deformation response, the sensitivity to a laser power as high as 0.07 nm/mW was obtained. The linear superposition behavior that validates the thermoelasticity theory was also confirmed, which guarantees surface error correction over a wide range of periodicities through laser spot scaling. It was also shown that the generated heat bump effectively compensates for residual shape errors in the mirror. Similar to bimorph mirrors, this kind of one-dimensional surface correction is sufficient for synchrotron radiation mirrors.

Future efforts using this technique may include the development of a new high-performance laser source to improve the laser-heat efficiency, along with application of this source to wavefront correction for synchrotron radiation in combination with at-wavelength metrology. The advantage of at-wavelength metrology over an interferometer is that the incidence of the laser can be normal and distortion of the projecting area can be avoided. This is very important for corrections with high spatial resolution. Further, the thermal field related to the final surface deformation differs for different cooling mechanisms (air convection and aluminum bars were used here), laser wavelengths, mirror materials, etc., which then affects the surface deformation shape. All of these factors should be considered in order to further develop the proposed technique. Moreover, this technique can be easily extended to high-resolution two-dimensional surface modulation through use of a spatial light modulator (such as a digital micromirror device (DMD) or liquid crystal display (LCD)).

**Funding.** National Science Foundation (NSF) (11505212); Jialin Xie Research Fund in the Institute of High Energy Physics (IHEPZZBS502).

## References

1. P. Emma, , R. Akre, , J. Arthur, , R. Bionta, , C. Bostedt, , J. Bozek, and Y. Ding, Nat. Photonics, **4**, 641 (2010).
2. ESRF Upgrade Programme, http://www.esrf.eu/about/upgrade
3. K. Yamauchi, H. Mimura, K. Inagaki, and Y. Mori, Rev. Sci. Instrum. **73**, 4028 (2002).
4. P. Revesz, A. Kazimirov, and I. Bazarov, Nucl. Instrum. Meth. A, **576**, 422 (2007).
5. S. Rutishauser, A. Rack, T. Weitkamp, Y. Kayser, C. David, and A. T. Macrander, J. Synchrotron Rad. **20**, 300 (2013).
6. H. Wang, K. Sawhney, S. Berujon, J. Sutter, S. G. Alcock, U. Wagner, and C. Rau, Opt. Lett., **39**, 2518 (2014).
7. H. Mimura, S. Handa, T. Kimura, H. Yumoto, D. Yamakawa, H. Yokoyama, S. Matsuyama, K. Inagaki, K. Yamamura, Y. Sano, K. Tamasaku, Y. Nishino, M. Yabashi, T. Ishikawa, and K. Yamauchi, Nat. Phys. **6**, 122 (2010).
8. M. Idir, P. Mercere, M. H. Modi, G. Dovillaire, X. Levecq, S. Bucourt, and P. Sauvageot, Nucl. Instrum. Meth. A **616**, 162 (2010).
9. Thales-SESO bimorph brochure, http://www.seso.com/
10. A. M. Morales, C. M. Lieber, Science **279**, 208 (1998).
11. J. Trice, C. Favazza, D. Thomas, H. Garcia, R. Kalyanaraman, and R. Sureshkumar, Phys. Rev. Lett. **101**, 017802. (2008).
12. M. E. Innocenzi, H. T. Yura, C. L. Fincher, and R. A. Fields, Appl. Phys. Lett. **56**, 1831 (1990)
13. F. Sato, L. C. Malacarne, P. R. B. Pedreira, M. P. Belancon, R. S. Mendes, M. L. Baesso, and J. Shen, J. Appl. Phys. **104**, 053520 (2008).
14. L.P.Welsh, J.A. Tuchman, I.P. Herman, J. Appl. Phys. **64**, 6274 (1988).
15. N.G.C. Astrath, F.B.G. Astrath, J. Shen, J. Zhou, P.R.B. Pedreira,L.C. Malacarne, A.C. Bento, M.L. Baesso, Opt. Lett. **33**, 1464(2008)
16. R. B. Hetnarski, *Encyclopedia of Thermal Stresses*, 1st ed. (Springer, 2014).